# Second- and third-order optical susceptibilities in bidimensional semiconductors near excitons states

Lucas Lafeta[1], Aurea Corradi[1], Tianyi Zhang[2,3], Ethan Kahn[2,3], Ismail Bilgin[4], Bruno R. Carvalho[5], Swastik Kar[4], Mauricio Terrones[2,3,6] and Leandro M. Malard[1]

[1] Departamento de Física, Universidade Federal de Minas Gerais, Belo Horizonte, Minas Gerais 30123-970, Brazil.
[2] Center for 2-Dimensional and Layered Materials, The Pennsylvania State University, University Park, PA 16802, USA.
[3] Department of Materials Science and Engineering, The Pennsylvania State University, University Park, PA 16802, USA.
[4] Department of Physics, Northeastern University, Boston, Massachusetts 02115, USA.
[5] Departamento de Física, Universidade Federal do Rio Grande do Norte, Natal, Rio Grande do Norte 59078-970, Brazil.
[6] Department of Physics, The Pennsylvania State University, University Park, PA 16802, USA.

E-mail: lmalard@fisica.ufmg.br

## Abstract

Semiconducting Transition Metal Dichalcogenides (TMDs) have significant nonlinear optical effects. In this work we have used second-harmonic generation (SHG) and the four-wave mixing (FWM) spectroscopy in resonance with the excitons in $MoS_2$, $MoSe_2$, and $WS_2$ monolayers to characterize the nonlinear optical properties of these materials. We show that trions and excitons are responsible for enhancing the nonlinear optical response, and determine the exciton and trion energies by comparing with the photoluminescence spectra. Moreover, we extract the second and third order optical sheet susceptibility ($\chi^{(2)}$ and $\chi^{(3)}$) near exciton energies and compare with values found in the literature. We also demonstrate the ability to generate different nonlinear effects in a wide spectral range in the visible region for monolayer $MoS_2$, opening the possibility of using two-dimensional materials for nonlinear optoelectronic and photonic applications.

## 1. Introduction

Enhancing light matter interactions is a key step towards the integration of two-dimensional (2D) materials in novel optoelectronic devices [1–4]. Two-dimensional semiconductors based on 2D transition metal dichalcogenides (TMDs) have strong many body effect interactions with high exciton binding energy, around hundreds of meV, due to a combination of quantum confinement and low-value dielectric constant [4,5], ultimately leading to an enhanced light absorption. These materials also exhibit substantial second- and third-order nonlinear optical responses [6,7], that have been theoretically explained due to the excitonic resonances [8–12]. In this context, we have probed the second and third-order optical process in different semiconducting 2D monolayer TMDs covering the first two excitonic levels, showing that excitonic states are crucial for understanding their unique nonlinear optical response.

Second-harmonic generation (SHG) has played a fundamental tool in the study of crystalline structure of different TMDs and its heterostructures [13–20]. It is a second-order process and allows us to obtain the values of the second order susceptibility, an intrinsic characteristic of the material, unveiling how responsive the material is to the second-order nonlinear phenomena as a function of the excitation laser energy. On the other hand, four-wave mixing (FWM) is a third-order nonlinear optical process and has been widely used for developing optoelectronic devices for generating a signal at shorter wavelengths originated from two signals at longer wavelengths [21–24]. Different works have explored the nonlinear response of different TMDs near to the excitonic energies [13,15,24–29]. However, a detailed spectroscopic study for the characterization of second- and third-order optical susceptibility ($\chi^{(2)}$ and $\chi^{(3)}$) in a wide range of energies across excitonic states are still lacking. Moreover, few works have shown the spectral characterization of the FWM response in these materials [24], and to the best of our knowledge, no more works have reported the FWM response for semiconducting monolayer TMDs.

In this work, we have studied the spectral response of second and third-order nonlinear response under resonance with the excitonic states in monolayers of $MoS_2$, $MoSe_2$, and $WS_2$. We found that the SHG and FWM signals are enhanced near excitonic energies, giving rise to well defined peaks in the spectra. By comparing with the photoluminescence spectra, we attribute the different resonance peaks to the A and B excitons, and trions states present in these materials. Moreover, the calibration of the SHG and FWM spectra allowed us to characterize the second- and third-order nonlinear sheet susceptibilities ($\chi^{(2)}$ and $\chi^{(3)}$) which is compared with other experimental data present in the literature. As an application of the high nonlinear optical susceptibilities found we show sum frequency generation (SFG), SHG and FWM operating simultaneously over a wide spectral range in a $MoS_2$ monolayer sample. Our work catalogs the second- and third-order susceptibilities in semiconducting monolayer TMDs near the excitonic levels, and shows the impact on the use of nonlinear optical techniques to study excitons in these materials, for characterization of the nonlinear properties and application in optoelectronic and photonic 2D devices [30–32].

## 2. Methods

Monolayers $MoSe_2$ were synthesized by chemical vapor deposition (CVD) at atmospheric pressure at 750°C, using selenium tablets (Se, Sigma-Aldrich, 99,9%) and powdered molybdenum oxide ($MoO_3$, Sigma-Aldrich, 99% ) [33]. For $MoS_2$ and $WS_2$, the samples were also grown by CVD but at 700°C. To synthesize $MoS_2$, molybdenum oxide ($MoO_3$, Sigma-Aldrich, 99%) was placed in an alumina boat, while for $WS_2$, tungsten oxide was used ($WO_3$, Alfa Aesar, 99.998%) [34]. All samples were grown on a $SiO_2/Si$ substrate. After synthesis, $MoSe_2$, $MoS_2$, and $WS_2$ were transferred to a fused quartz substrate, using poly (methyl methacrylate) (PMMA) [35] to avoid effects of interference generated by the $SiO_2/Si$ interface in our optical spectra [36].

Figure 1(a) shows the setup used in our experiments. For photoluminescence (PL) measurements, a 561 nm continuous wave (CW) diode laser, with 60 μW power, was used. Laser beam was reflected by the silver mirrors (M1 and M2), flip down the mirror flip (MF), and focus in the sample (S), placed on an inverted microscope (Nikon Eclipse), by the long-pass dichroic mirror (DC) and apochromatic objective (O) with 60 magnification, and 0.95 numerical aperture, giving a spot size of ~1 $\mu$m, backscattering signal is collected by the same objective and pass through the DC, reflected by mirrors (M4 and M5), filtered by the long-pass (LP) and focused by the lens in spectrometer (Andor Shamrock SR-303i-B) equipped with a charge coupled device (iDus DU420A-BEX2-DD).

For nonlinear measurements, a picosecond laser (APE PicoEMERALD) was used. This laser system can generate up to three laser beams that can be spatially and temporally overlapped with different wavelengths: at 1064 nm ($\lambda_{1064}$) with 7 ps pulse duration, tunable between 730 to 960 nm ($\lambda_{Signal}$) and between ~1193 to ~1961 nm ($\lambda_{Idler}$), both with 5-6 ps pulse duration. The $\lambda_{Signal}$ and $\lambda_{Idler}$ are generated by pumping the optical parametric oscillator cavity with the second harmonic of the 1064 nm laser at 532 nm. The selected $\lambda_{Signal}$, $\lambda_{Idler}$, and $\lambda_{1064}$ lasers are directed to the sample by the silver mirrors (M3 and MF), and collected the measure similar to PL, the only changes are changing dichroic mirror for beam-splitter (BS) and long-pass for short-pass (SP). The laser powers were kept at ~3 mW for $\lambda_{Signal}$, and $\lambda_{1064}$, and ~5 mW for $\lambda_{Idler}$.

Different nonlinear optical effects can be generated in monolayer TMDs when different lasers are focused onto the samples. In the case of SHG, the incident $\omega_{Idler}$ laser generates a $\omega_{SHG} = 2\omega_{Idler}$ response, as shown at Figure 1(b). By choosing an appropriate energy for $\omega_{Idler}$ laser, it is possible to enhance the SHG signal when the $2\omega_{Idler}$ matches the exciton energy of the TMD (see Figure 1(b), right side). In the case of FWM, three lasers focus on the material (two $\omega_{Signal}$ and one $\omega_{1064}$) and then combined to generate a new frequency at $\omega_{FWM} = 2\omega_{Signal} - \omega_{1064}$, see Figure 1(c). Similarly, the FWM signal is enhanced when this generated frequency is near the excitonic levels (see Fig. 1(c), right-side).

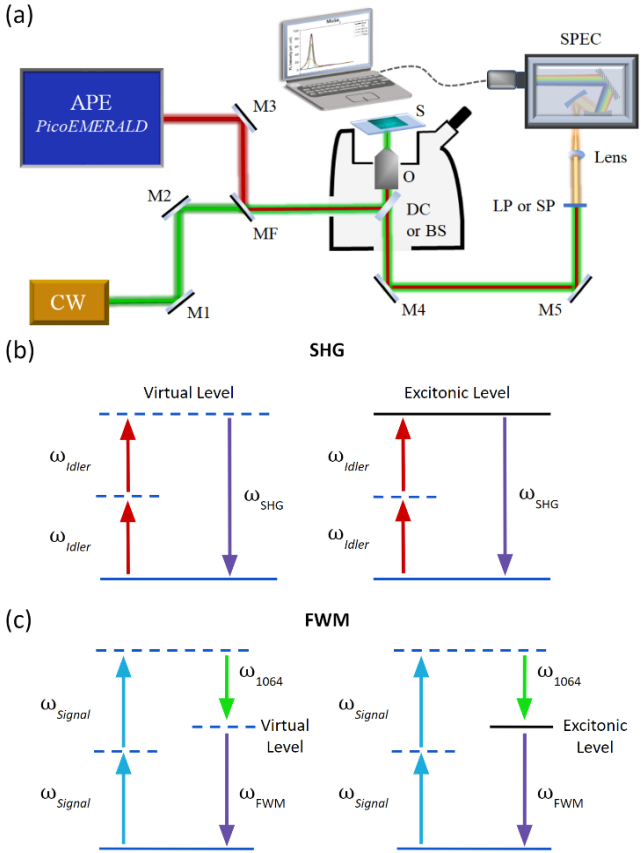

**Figure 1:** (a) Illustration for experimental setup used for the measurements. (b) Diagram of the SHG and (c) FWM energy transition off (left-side) and on (right-side) resonance near an exciton state.

## 3. Results and Discussion

Figures 2(a, b, c) show the collected PL spectra (black line) for $MoSe_2$, $MoS_2$ and $WS_2$ monolayers, respectively. The PL spectra were fitted by three Gaussian curves [37,38]: the trion (yellow line), exciton A (green line), and exciton B (blue line) that together generates the overall fitted curve (read line) for each TMD monolayer. Figures 2(d, e, f) show the SHG results for each sample. SHG measurements were performed by tuning the *Idler* laser beam ($2\omega_{Idler}$) to reach energies close to the A and B excitons of $MoSe_2$ and $MoS_2$, and the A exciton energy for the $WS_2$. Analogously, the FWM spectra were collected using the *Signal* and 1064 laser beams ($2\omega_{Signal} - \omega_{1064}$) are near the exciton energy levels, as shown in Figures 2(g, h, i). The SHG and FWM spectra were fitted by Lorentzians [39,40] curves for trion, and A and B excitons, using the same color pattern as for the PL spectra. The clear differentiation of each exciton peak, demonstrates that SHG and FWM spectroscopy can also be useful for the study and characterization of excitonic states in these 2D systems.

From Figure 2, one can observe that the PL peaks of the excitons and trions closely match the energies measured by SHG and FWM. The corresponding excitons and trion energy values extracted from the fitting procedure of the PL, SHG and FWM spectra are shown in Table-I. It is interesting to note the overall redshift of the PL exciton and trion energies compared to SHG and FWM in our measurements, especially for $MoS_2$. Since they are different physical processes, where PL depends on recombination mechanisms that are affected by phonon, temperature and defects while the nonlinear processes are not. Nonetheless our results closely match the results of the literature, even in materials deposited on different substrates and with different growth techniques (Table-I). The fact that the exciton binding energy is sensitive to the temperature, doping and defects present in the sample may be a possible explanation for the small differences between the results [38,41,42]. Moreover, our experimental results are in good agreement with the theoretical values reported [8–11].

It is worth noting that Figure 2 indicates that the SHG and FWM intensities increase several times under resonance conditions near the excitons and trion energies. This enhancement occurs because the non-linear process will occur through the interaction with a real state [43] as shown schematically in Figure 1(a, b). The presence of excitons or trions, therefore, makes the transition more likely to occur, as reported in previous works for SHG [13,15,25–28] and FWM [24].

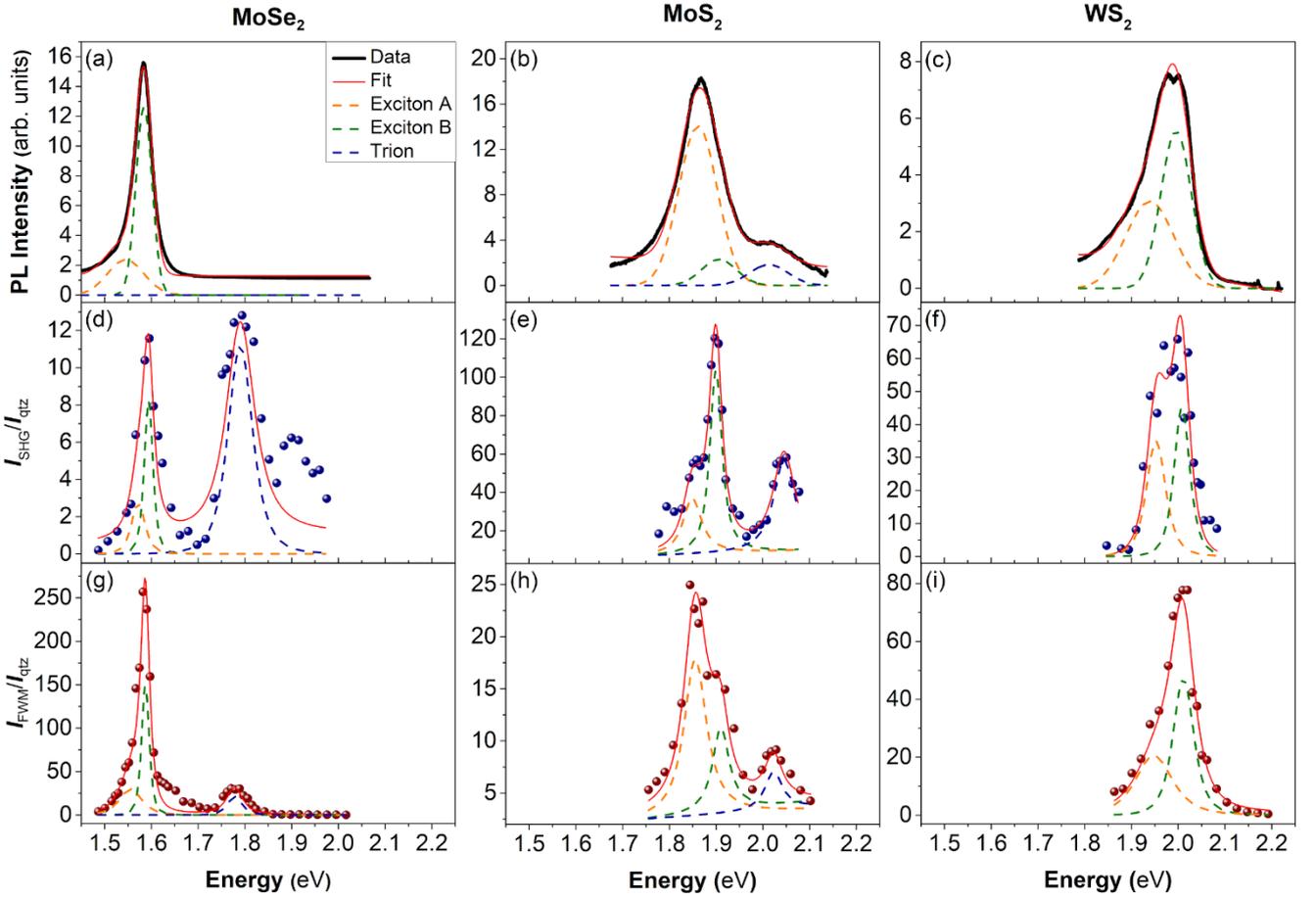

**Figure 2:** PL spectra for: (a) $MoSe_2$, (b) $MoS_2$ and (c) $WS_2$. SHG spectra for: (d) $MoSe_2$, (e) $MoS_2$ and (f) $WS_2$. FWM spectra for: (g) $MoSe_2$, (h) $MoS_2$ and (i) $WS_2$. The dashed green lines are the fitting curves for A exciton, blue for B exciton, yellow for trion, and red for the total fit.

**Table-I:** Energies values of A, B excitons and trions measured for semiconducting monolayer TMDs.

| TMD | Method | Trion (eV) | Exciton A (eV) | Exciton B (eV) | Ref. |
|---|---|---|---|---|---|
| $MoSe_2$ | | | | | |
| CVD | PL | 1.55 | 1.58 | 1.75 | this work |
| PLD | SHG | 1.50 | 1.54 | 1.77 | [27] |
| ME | PL | 1.55 | 1.58 | - | [44] |
| CVD | SHG | 1.57 | 1.59 | 1.79 | this work |
| CVD | FWM | 1.55 | 1.59 | 1.78 | this work |
| $MoS_2$ | | | | | |
| CVD | PL | 1.86 | 1.91 | 2.01 | this work |
| ME | PL | 1.87 | 1.91 | 2.06 | [44] |
| CVD | SHG | 1.85 | 1.90 | 2.04 | this work |
| CVD | SHG | - | 1.85 | 2.00 | [13] |
| ME | SHG | - | 1.91 | 2.07 | [25] |
| CVD | SHG | - | 1.87 | 2.07 | [26] |
| CVD | FWM | 1.86 | 1.91 | 2.02 | this work |
| ME | FWM | - | 1.82 | 2.07 | [24] |
| CVD | Ref. | - | 1.87 | 2.02 | [45] |
| $WS_2$ | | | | | |
| CVD | PL | 1.94 | 2.00 | - | this work |
| ME | PL | 1.98 | 2.01 | - | [44] |
| CVD | SHG | 1.96 | 2.01 | - | this work |
| CVD | FWM | 1.95 | 2.01 | - | this work |
| ME | Abs. | - | 2.02 | 2.40 | [46] |

PLD: Pulsed-laser-deposition; ME: Mechanical Exfoliation; Abs.: Absorption; Ref.: Reflectance.

To quantify the nonlinear susceptibility of the TMDs monolayers, the second- and third-order susceptibilities can be extracted from the SHG and FWM experimental results, shown in Fig. 1. Here, we compute the so-called nonlinear sheet susceptibility ($\chi^{(2)}_{TMD}$ and $\chi^{(3)}_{TMD}$) since the effect occurs on the 2D material's surface. To avoid any possible substrate's interference, we have collected the nonlinear spectrum of the quartz crystal for SHG and fused quartz for FWM under the same experimental condition of the samples. Therefore, the second- and third-order nonlinear susceptibilities sheet as a function of the excitation wavelength are given by [15, 47]:

$$\chi^{(2)}_{TMD} = \chi^{(2)}_{qtz} \left(\frac{I^{SHG}_{TMD}}{I^{SHG}_{qtz}}\right)^{\frac{1}{2}} \frac{(NA)^2 \pi c f}{4 \omega_{Idler} n}, \quad \text{(eq-1)}$$

$$\chi^{(3)}_{TMD} = \chi^{(3)}_{qtz} \left(\frac{I^{FWM}_{TMD}}{I^{FWM}_{qtz}}\right)^{\frac{1}{2}} \frac{c(1+n)^3}{16(2\omega_{signal} - \omega_{1064}) n^{\frac{5}{2}}}, \quad \text{(eq-2)}$$

where c is the speed of light in vacuum, $n = 1.45$ is the refractive index of quartz, $NA = 0.95$ is the objective's numerical aperture, $f = 0.056$ is a constant of integration, $\chi^{(2)}_{qtz} = 0.8 \times 10^{-12}$ mV$^{-1}$ is the second-order nonlinearity susceptibility for crystalline quartz [15], $\chi^{(3)} = 2 \times 10^{-22}$ m$^2$V$^{-2}$ is the third-order nonlinearity susceptibility for fused quartz [48], and $\omega_{Signal}$, $\omega_{1064}$, and $\omega_{Idler}$ are the excitation wavelength.

Figure 3 show the extracted $\chi^{(2)}_{TMD}$ and $\chi^{(3)}_{TMD}$ sheet for MoSe$_2$, MoS$_2$, and WS$_2$ monolayers as a function of the nonlinear laser (SHG and FWM) energy. The error bars are the standard deviation taken from an arithmetic mean of three different spectra.

In Table-II, we compare the values of $\chi^{(2)}$ found in our experiments with the values in the literature. It is possible to observe that the values obtained in the Refs. [10,15,17] for MoS$_2$ are in close agreement with our values. However, Refs. [16,19] differs in orders of magnitude. In the other materials, MoSe$_2$ and WS$_2$, our results are in close agreement with those from the literature, emphasizing that our experiment covers a wider spectral range, on and off resonance with the excitonic states, as shown in Fig. 3.

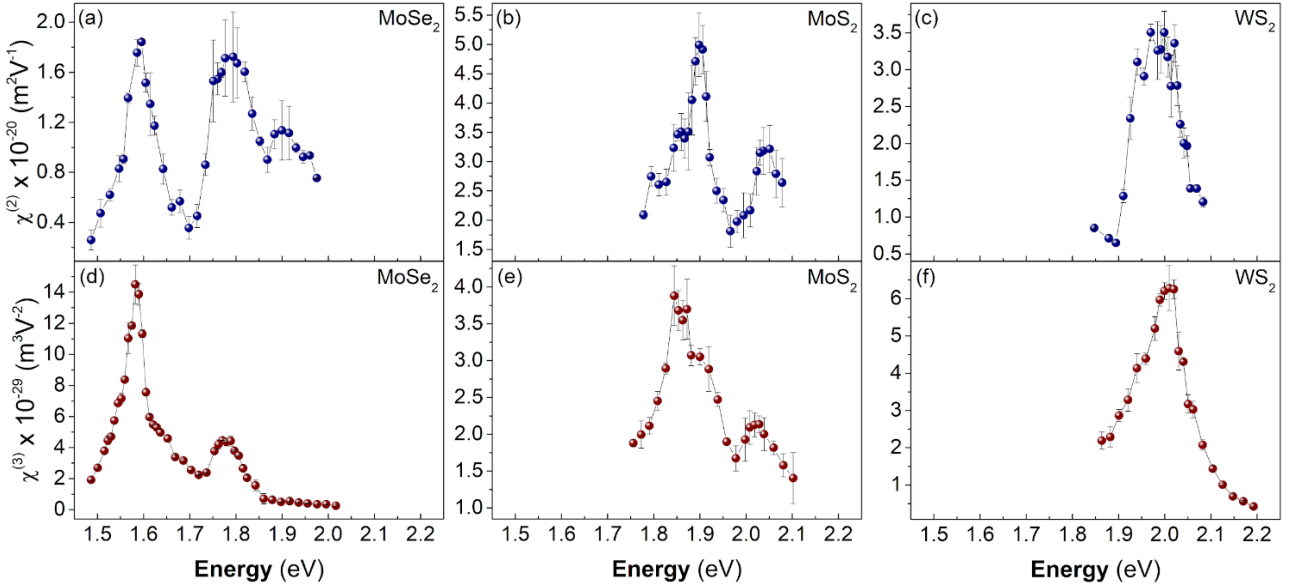

**Figure 3:** Sheet $\chi^{(2)}$, close to the exciton energies of (a) MoS$_2$, (b) MoSe$_2$, and (c) WS$_2$, and sheet $\chi^{(3)}$, close to the exciton energies of (a) MoS$_2$, (b) MoSe$_2$, and (c) WS$_2$.

**Table-II:** Extracted sheet $\chi^{(2)}$ values from our experiments and from results from the literature with its corresponding excitation laser wavelength (λ laser) and the substrate where the sample is deposited.

| TMD | $\chi^{(2)}$ times $10^{-20}$ (m$^2$V$^{-1}$) | Method | λ laser (μm) | Substrate | Ref. |
|---|---|---|---|---|---|
| MoSe$_2$ | | | | | |
| CVD | 0.2-2.3 | SHG | 1.26-1.67 | SiO$_2$ | this work |
| ME | 2.4 | SHG | 1.56 | Si/SiO$_2$ | [10] |
| PLD | 0.6-4.0 | SHG | 1.2-1.8 | Si/SiO$_2$ | [27] |
| MoS$_2$ | | | | | |
| CVD | 1.7-5.1 | SHG | 1.19-1.40 | SiO$_2$ | this work |
| CVD | 0.3-2.9 | SHG | 1-1.8 | Si/SiO$_2$ | [13] |
| CVD | 3.3 x 10$^2$ | SHG | 0.81 | SiO$_2$ | [16] |
| CVD | 2.0 | SHG | 1.56 | glass | [17] |
| CVD | 0.4-4.2 | SHG | 1.06-1.6 | SiO$_2$ | [26] |
| ME | 1.1-8.0 | SHG | 0.68-1.08 | quartz | [15] |
| ME | 6.5 x 10$^3$ | SHG | 0.81 | SiO$_2$ | [16] |
| ME | 2.0 x 10$^3$ | SHG | 0.81 | glass | [19] |
| ME | 0.35 | SHG | 1.56 | SiO$_2$ | [10] |
| WS$_2$ | | | | | |
| CVD | 1.2-4.2 | SHG | 1.2-1.24 | SiO$_2$ | this work |
| CVD | 5.9 | SHG | 0.832 | Si/SiO$_2$ | [18] |
| ME | 1.1 | SHG | 1.56 | Si/SiO$_2$ | [10] |

In Table-III, we compared the values of the sheet $\chi^{(3)}$ obtained in Fig. 3 and with the values reported in the literature. It is possible to observe that our values are within the same order of magnitude compared to the ones found in the literature by third harmonic generation (THG). Since THG is done with lasers energies where the generated light has energies above the exciton and band gap energies, the values of sheet $\chi^{(3)}$ in the literature are larger compared to ours.

**Table-III:** Extracted sheet $\chi^{(3)}$ values from our experiments and from results from the literature with its corresponding excitation laser wavelength (λ laser) and the substrate where the sample is deposited.

| TMD | $\chi^{(3)}$ times $10^{-29}$ (m$^2$V$^{-1}$) | Method | λ laser (nm) | Substrate | Ref. |
|---|---|---|---|---|---|
| MoSe$_2$ | | | | | |
| CVD | 0.1-14.7 | FWM | 780-820* | SiO$_2$ | this work |
| ME | 14.3 | THG | 1560 | Si/SiO$_2$ | [10] |
| MoS$_2$ | | | | | |
| CVD | 1.3-4.0 | FWM | 760-850* | SiO$_2$ | this work |
| CVD | 17 | THG | 1560 | glass | [17] |
| CVD | 7.8 | THG | 1560 | SiO$_2$ | [49] |
| ME | 23.4 | THG | 1560 | Si/SiO$_2$ | [10] |
| ME | 6.5 | THG | 1758 | Si/SiO$_2$ | [50] |
| WS$_2$ | | | | | |
| CVD | 0.2-6.5 | FWM | 740-820* | SiO$_2$ | this work |
| ME | 15.6 | THG | 1560 | Si/SiO$_2$ | [10] |

* The λ values of the laser refers to the Signal laser, whereas the 1064 nm laser was kept fixed.

The large second and third order susceptibilities found here near the excitons energies opens up the opportunity to generate nonlinear optical effects in a large spectral range in these materials. By focusing three different lasers ($\omega_{Signal}$, $\omega_{1064}$, and $\omega_{Idler}$) simultaneously in a monolayer MoS$_2$ sample on a Si/SiO$_2$ substrate is possible to generate different nonlinear optical effects such as SHG (2$\omega_{Signal}$, 2$\omega_{1064}$, and 2$\omega_{Idler}$), SFG ($\omega_{Signal}$ + $\omega_{1064}$, $\omega_{Signal}$ + $\omega_{Idler}$, and $\omega_{1064}$ + $\omega_{Idler}$), and FWM (2$\omega_{Signal}$ - $\omega_{Idler}$ and 2$\omega_{Signal}$ - $\omega_{1064}$). All of these processes are schematically shown in Figure 4(a). In Fig. 4(b) it is possible to observe these different nonlinear effects in almost the entire region of the visible spectrum (400 to 720 nm) by varying the Signal laser in 10 nm steps.

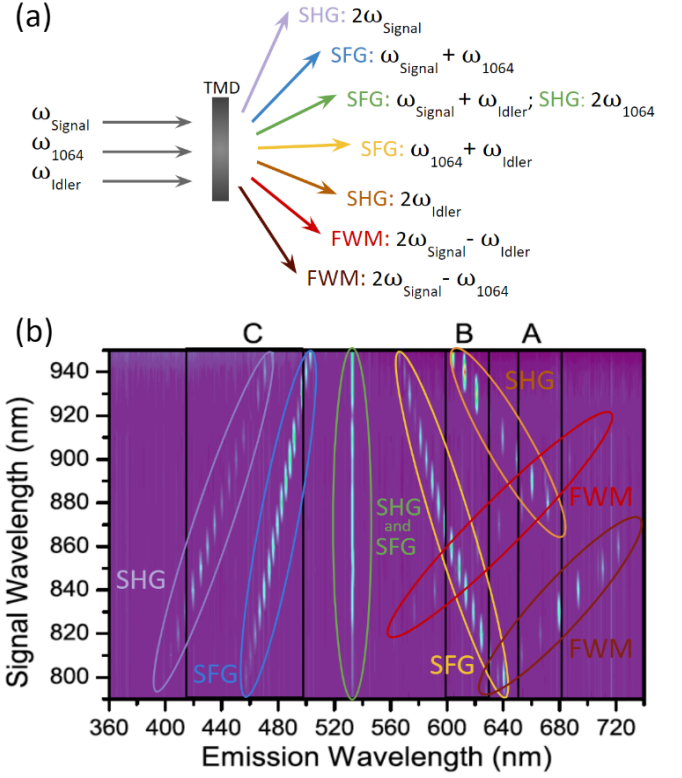

**Figure 4:** (a) Illustration of different nonlinear processes due to the interaction of the different lasers ($\omega_{Signal}$, $\omega_{Idler}$, and $\omega_{1064}$ lasers) on the MoS$_2$ sample. (b) Intensity map of nonlinear effects varying the wavelength of the $\omega_{Signal}$ laser. The regions circled with different colors refer to the colors of nonlinear processes indicated in part (a). The black rectangles indicate the energy positions of the A, B and C excitons.

## 4. Conclusions

In conclusion, we have shown that SHG and FWM allowed us to study excitons and trions states in different TMDs. We have observed an increased SHG and FWM intensities close to the exciton and trion energies, showing that the TMDs have a high nonlinear optical response. Moreover, it was possible to extract the second and third-order susceptibility values for the different materials near the exciton energies. Due to the increase in second and third-order susceptibilities, we have also shown that different non-linear effects are possible over a wide spectral range for MoS$_2$. Our results could be useful for future comparison of theoretical calculation of the nonlinear optical susceptibilities in these materials. Also, our nonlinear optical characterization could be useful in designing future nonlinear optoelectronic devices based on two-dimensional materials.

**Acknowledgments**

We acknowledge the Brazilian funding agencies CNPq, FAPEMIG, CAPES, FINEP, the Brazilian Institute of Science and Technology (INCT) in Carbon Nanomaterials and in Molecular Medicine for the financial support. T.Z. and M.T. also thank the financial support from the National Science Foundation (2DARE-EFRI-1433311). S.K. acknowledges financial support from NSF ECCS 1351424.